\documentstyle[prl,preprint,aps]{revtex}

\begin{document}
\draft
\preprint{HEP/123-qed}
\title{Bounds for the time to failure of hierarchical systems of fracture}
\author{J. B. G\'{o}mez}
\address{Departamento de Ciencias de la Tierra,
 Universidad de Zaragoza, 50009 Zaragoza, Spain.}
\author{M. V\'{a}zquez-Prada, Y. Moreno\cite{byline}, and A. F. Pacheco}
\address{Departamento de F\'{\i}sica Te\'{o}rica, Universidad de Zaragoza, 50009 Zaragoza,
Spain.}
\date{\today}
\maketitle
%\widetext
\begin{abstract}
For years limited Monte Carlo simulations have led to the suspicion
that the time to failure of hierarchically organized load-transfer
models of fracture is non-zero for sets of infinite size. This fact
could have a profound significance in engineering practice and also in
geophysics. Here, we develop an exact algebraic iterative method to
compute the successive time intervals for individual breaking in
systems of height $n$ in terms of the information calculated in the
previous height $n-1$. As a byproduct of this method, rigorous lower
and higher bounds for the time to failure of very large systems are
easily obtained. The asymptotic behavior of the resulting lower bound
leads to the evidence that the above mentioned suspicion is actually
true.
\end{abstract}
%\begin{multicols}{2}
\narrowtext

%\section{First-level heading:}
%\label{sec:level1}

There are few mechanical problems more complex and difficult to cast
into a definite physical and theoretical treatment than the range of
phenomena associated with fracture. Furthermore, there are few
problems with a wider field of application: material science,
engineering, rock mechanics, seismology and earthquake occurrence. Our
understanding of fracture processes in heterogeneous materials has
recently improved with the development of simple deterministic and
stochastic algorithms to simulate the process of quasistatic loading
and static fatigue  \cite{herrman90}. The load-transfer models used
here are called fiber-bundle models, because they arose in close
connection with the strength of bundles of textile fibers. Since
Daniels' and Coleman's \cite{daniels45} seminal works, there has been
a long tradition in the use of these models to analyze failure of
heterogeneous materials \cite{harlow78}. Fiber-bundle models differ by
how the load carried by a failed element is distributed among the
surviving elements in the set. In the simplest scheme, called ELS for
equal load sharing, the load supported by failed elements is shared
equally among all surviving elements. Important from the geophysical
point of view is the hierarchical load-sharing (HLS) rule introduced
by Turcotte and collaborators in the seismological literature
\cite{smalley85}. In this load-transfer modality the scale invariance
of the fracture process is directly taken into account by means of a
scheme of load transfer following the branches of a fractal (Cayley)
tree with a fixed coordination number $c$. The HLS geometry nicely
simulates the Green's function associated with the elastic
redistribution of stress adjacent to a rupture. In the static case,
the strength of an HLS system tends to zero as the size of the system,
$N$, approaches infinity, though very slowly, as
\( 1/log(log N)\) \cite{newman91}. The dynamic HLS model was introduced
in the geophysical literature in reference \cite{newman94}; their
specific aim was to find out if the chain of partial failure events
preceding the total failure of the set resemble a log-periodic
sequence \cite{sornette95}. In the analysis of \cite{newman94}, it
appeared that, contrary to the static model, the dynamic HLS model
seemed to have a non-zero time-to-failure as the size of the system
tends to infinity. In this paper we provide evidence that this
conjecture is true by means of an exact algebraic iterative method
where trees of
\(n\) levels, $N=c^n$, are solved using the information acquired in the previous
calculation of trees of \(n-1\) levels. As a byproduct of this method,
we obtain a rigorous lower bound for the time to failure of infinite
size sets which results in being non-zero.

In \cite{gomez98} we showed how in fiber-bundle dynamical models,
using a Weibull exponential shape function and a power-law breaking
rule, one can devise a probabilistic approach which is equivalent to
the usual approach \cite{newman94} where the random lifetimes are
fixed at the beginning and the process evolves deterministically.
Without loss of generality, from the probabilistic perspective the set
of elements with initial individual loads \(\sigma=\sigma_0=1\) and
suffering successive casualties is equivalent to an  inhomogeneous
sample of radioactive nuclei each with a decay width
\(\sigma^{\rho}\); \(\rho\) being the so called Weibull index, which
in materials science adopts typical values between 2 and 10. As time
passes and failures occur, loads are transferred,
\(\sigma\) changes and the decay width of the surviving elements grow.
In the probabilistic approach, in each time step defined as:
\begin{equation}
\delta=\frac{1}{\sum\limits_{j}\sigma_{j}^{\rho}}\quad ,
\label{one}
\end{equation}
one element of the sample decays. The index \(j\) runs along all the
surviving elements. The probability of one specific element,
\(m\), to fail is \(p_{m}=\sigma_{m}^{\rho} \delta\). Eq.\ (\ref{one}) is the ordinary
link between the mean time interval for one element to decay in a
radioactive sample and the total decay width of the sample as defined
above. Due to this analogy, radioactivity terms will frequently be
used in this communication. Note that we use loads and times without
dimensions. The time to failure, \(T\), of a set is the sum of the
\(N\)
\(\delta\)s. For the ELS case, the value of \(\delta\) as defined in
Eq.\ (\ref{one}) easily leads to $T=1/\rho$, which is the correct
result for the time to failure of an infinite ELS set \cite{newman94}.
In this communication we will apply these ideas to the HLS case
obtaining the
\(\delta\)s algebraically without having to use Monte Carlo
simulations.

To give a perspective of what is going on in the rupture process of a
hierarchical set we have drawn in Fig.\ \ref{figure1} the three
smallest cases for trees of coordination \(c=2\). Denoting by
\(n\) the number of levels, or height of the tree, i.e.
\(N=2^{n}\), we have considered \(n=0,1\) and
\(2\). The integers within parenthesis
\((r)\) account for the number of failures existing in the tree. When
there are several non-equivalent configurations corresponding to a
given
\(r\), they are labeled as \((r,s)\), i.e, we add a new index $s$.
The total load is conserved except at the end, when the tree
collapses. Referring to the high symmetry of loaded fractal trees,
note that each of the configurations explicitly drawn in Fig.\
\ref{figure1} represents all those that can be brought to coincidence
by the permutation of two legs joined at an apex, at any level in the
height hierarchy. Hence we call them non-equivalent configurations or
merely configurations. In general, each configuration $(r,s)$ is
characterized by its probability $p(r,s)$, $\sum\limits_{s}p(r,s)
=1$, and its decay width $\Gamma(r,s)$. The time step for one-element
breaking at the stage $r$, is given by
\begin{equation}
\delta_{r}=\sum\limits_{s}p(r,s)\frac{1}{\Gamma(r,s)} \label{two}
\end{equation}
This is the necessary generalization of Eq.\ (\ref{one}) due to the
appearance of non-equivalent configurations for the same $r$ during
the decay process of the tree. In cases of branching, the probability
that a configuration chooses a specific direction is equal to the
ratio between the partial decay width in that direction and the total
width of the parent configuration. We will compute at a glance the
\(\delta\)s of Fig.\ \ref{figure1} in order to later analyze the general
case. To be specific, we will always use \(\rho=2\). For
\(n=0\), we have $\Gamma(0)=1^{2}$ and \(\delta_{0}=\frac{1}{1^{2}}=1=T\).
For \(n=1\), \(\Gamma(0)=1^{2}+1^{2}=2\), $\delta_{0}=\frac{1}{2}$,
\(\Gamma(1)=2^{2}\), $\delta_{1}=\frac{1}{4}$ and hence \(T=\frac{1}{2}+\frac{1}{4}=\frac{3}{4}\). For \(n=2\),
\(\Gamma(0)=1^{2}+1^{2}+1^{2}+1^{2}=4\), \(\delta_{0}=\frac{1}{4}\),
\(\Gamma(1)=2^{2}+1^{2}+1^{2}=6\), $\delta_{1}=\frac{1}{6}$; now we face a branching,
the probability of the transition $(1)\rightarrow(2,1)$ is
$\frac{4}{6}$ and the probability of the transition
$(1)\rightarrow(2,2)$ is $\frac{2}{6}$, on the other hand
$\Gamma(2,1)=2^{2}+2^{2}=8=\Gamma(2,2)$, hence
$\delta_{2}=\frac{4}{6}\cdot\frac{1}{8}+\frac{2}{6}\cdot\frac{1}{8}=\frac{1}{8}$.
Finally $\delta_{3}=\frac{1}{16}$ and the addition of $\delta$s gives
$T=\frac{29}{48}$.

Now we define the replica of a configuration belonging to a given
\(n\), as the same configuration but with the loads doubled (this is
because we are using \(c=2\)). The replica of a given configuration
will be recognized by a prime sign. In other words $(r,s)'$ is the
replica of $(r,s)$. Any decay width, partial or total, related to
$(r,s)'$ is automatically obtained by multiplying the corresponding
value of $(r,s)$ by the common factor $c^{\rho}=2^{\rho}=4$. This is a
consequence of the power-law breaking rule assumed. The need to define
the replicas stems from the observation that any configuration
appearing in a stage of breaking $r$ of a given
\(n\), can be built as the juxtaposition of two configurations of
the level
\(n-1\), including also the replicas of the level $n-1$ as ingredients of the
game. In Fig.\ \ref{figure1}, one can observe the explicit structure
of the configurations of $N=4$ as a juxtaposition of those of $N=2$
and its replicas. From this perspective we notice that as the
configurations for the height $n$ are formed by juxtaposing two
configurations of the already solved height $n-1$ and its replicas,
with the restriction that the total load must be equal to $2^n$
because the total load is conserved, the single-element breaking
transitions occurring can only be of three types. One case a)
corresponds to the breaking of one element in a half of the tree while
the other half remains as an unaffected spectator. Another case b)
corresponds to the decay of the last surviving element in a half of
the tree, which provokes its collapse and the corresponding doubling
of the load borne by the other half. Finally, the third case c)
corresponds to the scenario in which one half of the tree has already
collapsed and in the other half one break occurs. In this third case
the decay width is obtained from the information of the replicas. This
holds for any height $n$, allows the computation of all the partial
decay widths in a tree of height $n$ from those obtained in the height
$n-1$ and will be illustrated in Fig.\ \ref{figure3}.

Using henceforth a convenient symbolic notation, in Fig.\
\ref{figure2}a we have built what we call the primary width diagram
for $n=3$ resulting from the juxtapositions of pairs of configurations
of the previous diagram of $n=2$ and its replicas. It is implicitly
understood that time flows downward along with the breaking process.
In Fig.\ \ref{figure2}a the boxes represent $n=3$ configurations,  and
at their right, the two quantities in parentheses indicate the two
$n=2$ configurations forming them. The value of the partial width
connecting a parent and a daughter is written at the end of the
corresponding arrow. The sum of all the partial widths of a parent
configuration in a branching is always equal to the total decay width,
$\Gamma$, of the parent. From this primary width diagram one deduces
the probability of any primary configuration at any stage $r$ of
breaking, and consequently $\delta_{r}$ is obtained using Eq.\
(\ref{two}). Finally, by adding all the $\delta$s we calculate
$T(n=3)$. To illustrate the connection between the explicit
configurations as those of Fig.\ \ref{figure1} and the somewhat
hermetic notation of Fig.\ \ref{figure2}, in Fig.\ \ref{figure3}, we
have shown three explicit examples relating them. Fig.\ \ref{figure3}
is selfexplanatory. The three examples correspond to the cases a), b)
and c) mentioned before.

By iterating this procedure, that is by forming the primary diagram of
the $n+1$ height by juxtaposing the configurations of the primary
diagram of the height $n$, we can, in principle, exactly obtain the
total time to failure of trees of successively doubled size. Two
examples are $T(n=3)=\frac{63451}{123200}$ and
$T(n=4)=\frac{21216889046182831}{46300977698976000}$. The problem
arises from the vast amount of configurations one has to handle in the
successive primary diagrams. This fact eventually blocks the
possibility of obtaining exact results for trees high enough as to be
able to gauge the asymptotic behavior of $T$ in HLS sets. That is why,
taking advantage of the general perspective gained with the exact
algebraic method, henceforth our more modest goal will be to set
rigorous bounds for $T$. This task is much simpler.

To set bounds, we will define effective diagrams in which for each $r$
there is only one configuration which results from fusing in some
specific appropriate way all the configurations labeled by the
different $s$ values; see Fig.\ \ref{figure2}b. These effective
configurations are connected by effective decay widths denoted by
$a_r$. Thus, the effective diagram for any $n$ is calculated from its
primary diagram, and then is used to build a primary diagram of the
next height $n+1$. For $n=0$, 1 and 2, $a_r=\Gamma(r)$ because
$\forall r$, the $\Gamma(r,s)$ are equal to a unique value
$\Gamma(r)$. The point is how to define $a_r$ for $n\ge 3$, so that
all the $\delta(r)(n\ge4)$ and hence $T(n\ge4)$ are systematically
lower (or higher) than its exact result. This goal is easily
accomplished using,
\begin{equation}
a_r=\Gamma_{max}(r)\qquad (or\quad \Gamma_{min}(r)) \label{three}
\end{equation}
i.e., by assuming that the configuration of maximum (minimum) width
saturates the single element decay of the stage $r$. Better rigorous
boundings are obtained using the geometric mean (or the harmonic
mean), namely
\begin{equation}
a_r=\prod\limits_{s}\Gamma(r,s)^{p(r,s)}\qquad (or\quad
\frac{1}{\sum\limits_{s}p(r,s)\frac{1}{\Gamma(r,s)}})
\label{four}
\end{equation}
The bounds obtained from these formulae are plotted in Fig.\
\ref{figure4}. It is clear that those based on Eq.\ (\ref{three}) are
poor, in fact the lower bound goes quickly to zero. On the other hand,
those based on Eq.\ (\ref{four}) are stringent and useful. The value
obtained for the lower (higher) bound to $T$, from Eq.\ (\ref{four}),
will be called $T_l$ $(T_h)$. The arithmetic mean, i.e.,
$a_r=\sum\limits_{s}p(r,s)\cdot\Gamma(r,s)$ also leads to a lower
bound but it is not as good as that coming from the geometric mean.
The bounds emerging from these three means are rigorous because when
configurations of different decay width are fused, the form of the
generated function appearing in the calculus of the $\delta$s are
concave (convex). This will be explained elsewhere.

We have fitted the data points of $T_l$ by an exponential function of
the form \(T_{l}=T_{l,\infty}+a e^{-b(n-n_{0})}\),
\(T_{l,\infty}\), \(a\), \(b\) and \(n_{0}\) being fitting parameters.
The success of this fitting is crucial because this exponential decay
to a non-zero limit is the hallmark of our claim. By choosing subsets
formed only by points representing big systems we observe a clear
saturation of the asymptotic time-to-failure towards
\(T_{l,\infty}=0.32537\pm0.00001\). An analogous analysis of $T_h$
leads to $T_{h,\infty}=0.33984\pm0.00001$. Similar fittings of the
Monte Carlo data points are inconclusive, due to the intrinsic
noisiness of the MC results and the limited size of the simulated sets
($N<2^{16}$ elements). What this result implies is that a system with
a hierarchical scheme of load transfer and a power-law breaking rule
$(c=2,\rho=2)$ has a time-to-failure for sets of infinite size,
$T_{\infty}$, such that $0.32537 \leq T_{\infty} \leq 0.33984$. Thus,
there is an associated zero-probability of failing for $T<T_{\infty}$
and a probability equal to one of failing for
\(T>T_{\infty}\). The critical point behavior is thus confirmed.
Invoking conventional universality arguments one deduces that this
property holds for a hierarchical structure of any coordination. The
case of dynamical HLS sets using an exponential breaking rule will be
reported shortly.

A.F.P is grateful to J. As\'{\i}n, J.M. Carnicer and L. Moral for
clarifying discussions. M.V-P. thanks Diego V\'{a}zquez-Prada for
discussions. Y.M thanks the AECI for financial support. This work was
supported in part by the Spanish DGICYT.

\begin{figure}
\caption{Breaking process for the three smallest trees of coordination $c=2$ ($N=1, 2, 4$).
 The integers in parenthesis $(r)$ represent the number of breakings occurred.
 The $\delta$s stand for the time steps elapsed between successive individual
 breakings and the numbers under the legs indicate the load they bear.}
\label{figure1}
\end{figure}

\begin{figure}
\caption{Width diagrams for the breaking process of $N=8$. a) Primary diagram.
The various configurations are labeled by the integers inside the
boxes. The integers in parentheses at the right of the boxes represent
the two $N=4$ configurations juxtaposed to form each of the $N=8$
configurations. The number accompanying an arrow connecting two boxes
stands for the dimensionless partial decay width of that transition.
b) Effective diagram. This is obtained from a), and is used to obtain
rigorous bounds for the next height, $N=16$. The $a$s are explained in
the text.}
\label{figure2}
\end{figure}

\begin{figure}
\caption{Relation between the explicit and the symbolic notations for
configurations of $N=8$, built juxtaposing pairs of $N=4$. See the
text for details.}
\label{figure3}
\end{figure}

\begin{figure}
\caption{Dimensionless lifetime, $T$, for a fractal tree of height $n$: the circles
 are obtained from Monte Carlo simulations; the continuous lines are the
 stringent bounds, and the dotted lines are the gross bounds.}
\label{figure4}
\end{figure}

%\end{multicols}
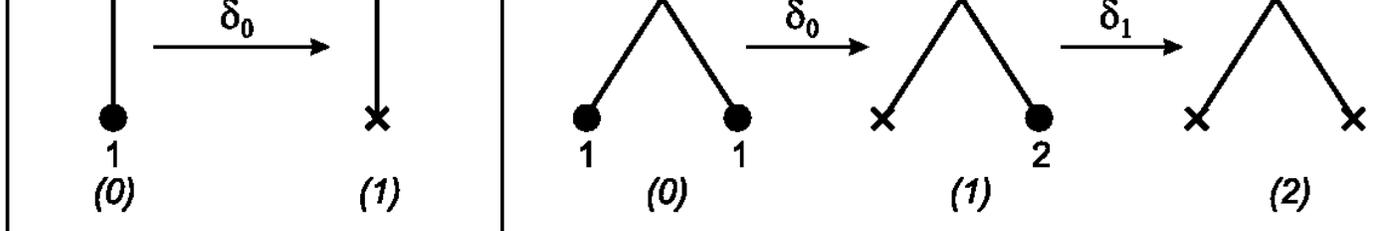
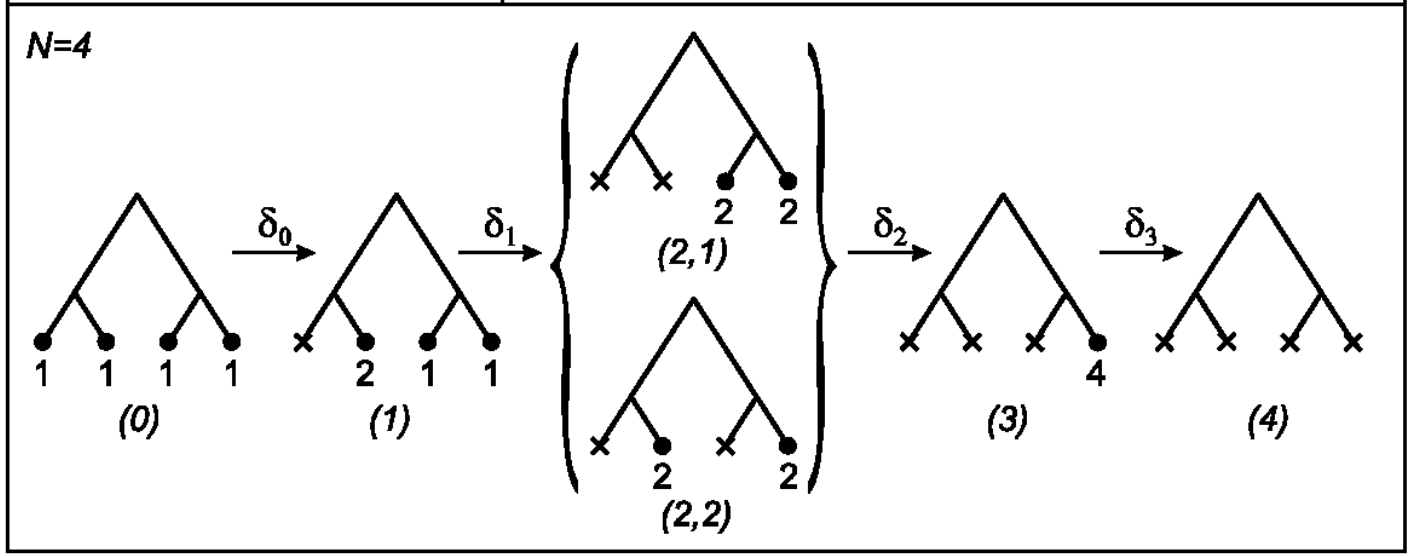
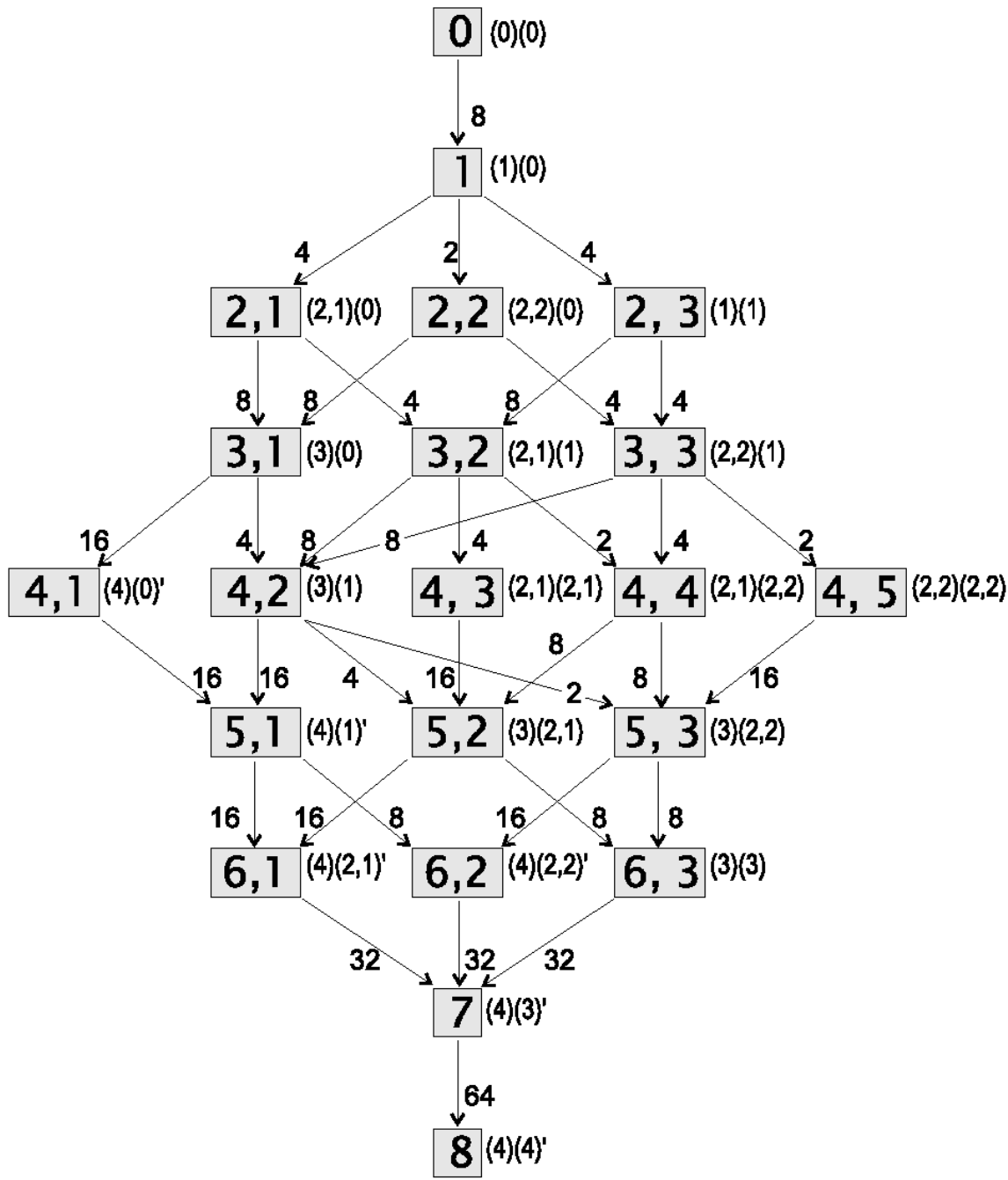
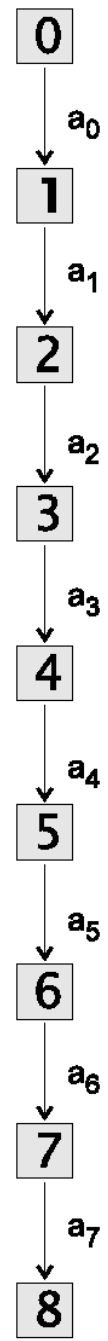
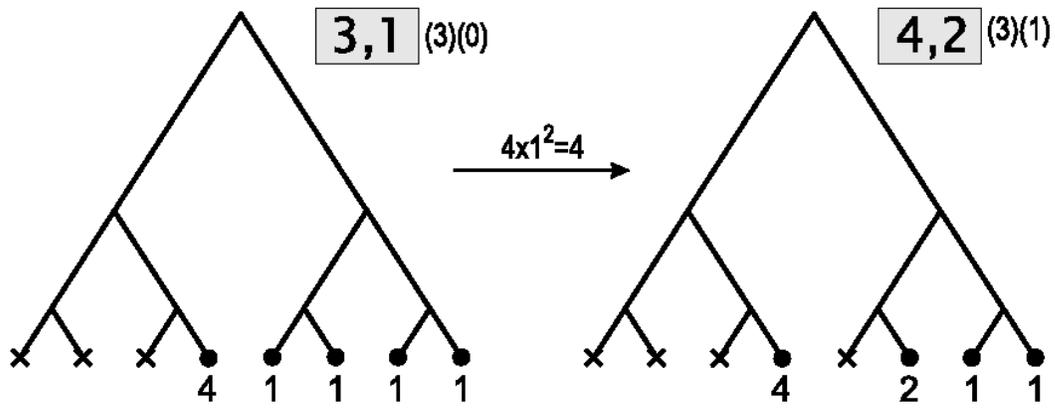
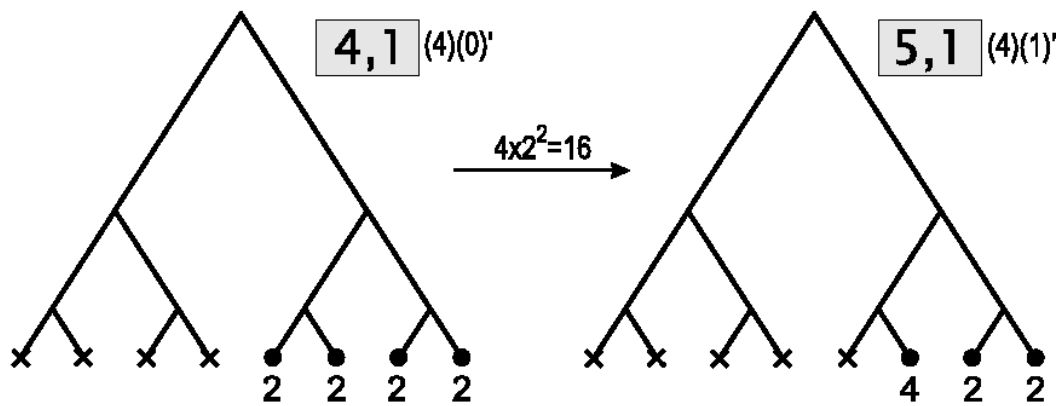
\end{document}